\documentstyle[11pt,epsf]{article}
\input psfig
\input AMSsymb.sty
\def\beq{\begin{eqnarray}}
\def\eeq{\end{eqnarray}}
\def\bea{\begin{eqnarray*}}
\def\eea{\end{eqnarray*}}

\def\centeron#1#2{{\setbox0=\hbox{#1}\setbox1=\hbox{#2}\ifdim
\wd1>\wd0\kern.5\wd1\kern-.5\wd0\fi
\copy0\kern-.5\wd0\kern-.5\wd1\copy1\ifdim\wd0>\wd1
\kern.5\wd0\kern-.5\wd1\fi}}
\def\ltap{\;\centeron{\raise.35ex\hbox{$<$}}{\lower.65ex\hbox{$\sim$}}\;}
\def\gtap{\;\centeron{\raise.35ex\hbox{$>$}}{\lower.65ex\hbox{$\sim$}}\;}

\def\singleandthirdspaced{\baselineskip=\normalbaselineskip\multiply
    \baselineskip by 130\divide\baselineskip by 100}

\def\dslash{\not{\hbox{\kern-2pt $\partial$}}}
\def\Dslash{\not{\hbox{\kern-4pt $D$}}}
\def\Oslash{\not{\hbox{\kern-4pt $O$}}}
\def\Qslash{\not{\hbox{\kern-4pt $Q$}}}
\def\pslash{\not{\hbox{\kern-2.3pt $p$}}}
\def\kslash{\not{\hbox{\kern-2.3pt $k$}}}
\def\qslash{\not{\hbox{\kern-2.3pt $q$}}}

\def\p0plus{(p^0+m)}
\def\p0minus{p^0-m-eA^0}

\newcommand{\newc}{\newcommand}
\newc{\qbar}{{\overline q}}
\newc{\Kahler}{K\"ahler }
\newc{\deltaGS}{\delta_{\rm GS}}
\begin{document}
\begin{titlepage}
\begin{flushright}
{\large hep-th/yymmnn \\ SCIPP-2010/04\\
}
\end{flushright}

\vskip 1.2cm

\begin{center}

{\LARGE\bf Reliable Semiclassical Computations in QCD}

\vskip 1.4cm

{\large Michael Dine$^{a,b}$, Guido Festuccia$^{a}$, Lawrence Pack$^{a}$ and Weitao Wu$^{a}$}
\\
\vskip 0.4cm
{\it $^a$Santa Cruz Institute for Particle Physics and
\\ Department of Physics, University of California,
     Santa Cruz CA 95064  } \\
{\it $^b$Department of Physics, Stanford University
Stanford, CA 94305-4060, USA}\\
\vskip 4pt

\vskip 1.5cm

\begin{abstract}
We revisit the question of whether or not one can perform
reliable semiclassical QCD computations at zero temperature.  We study correlation functions
with no perturbative contributions, and organize the problem
by means of the operator product expansion, establishing a precise criterion for the validity of a semiclassical calculation.
For $N_f>N$, a systematic computation is possible; for $N_f<N$, it is not.   $N_f=N$
is a borderline case.    In our analysis, we see explicitly the exponential
suppression of instanton effects at large $N$.  As an application, we describe a test
of QCD lattice gauge theory computations in the chiral limit.
\end{abstract}

\end{center}

\vskip 1.0 cm

\end{titlepage}
\setcounter{footnote}{0} \setcounter{page}{2}
\setcounter{section}{0} \setcounter{subsection}{0}
\setcounter{subsubsection}{0}

\singleandthirdspaced

\section{Introduction:  Perturbative and Non-Perturbative contributions to the OPE}

Over the last few decades, semiclassical computations have proven valuable in understanding a variety of effects in quantum
field theory and string theory.  In four dimensional gauge theories, 't Hooft showed in his original work that systematic computations
are possible in a Higgs phase\cite{thooft}.  It was soon realized that in QCD at high temperatures, instantons provide the leading
contribution to $\theta$-dependent quantities, as well as to certain chirality violating amplitudes\cite{gpy,affleck}.  These methods found applications
to the understanding of baryon number violation in the Standard Model, the
axion potential at high temperatures and the dynamics of supersymmetric gauge theories.
But the question of the utility of instantons at zero temperature in real QCD has remained elusive.  Even for short distance quantities, such computations are plagued
by infrared divergences.  These divergences arise precisely in the region of strong coupling, suggesting that instantons are, at best,
just one of many important non-perturbative effects.  Witten, indeed, argued compellingly that at large $N$, instanton effects are negligible\cite{wittencpn}.

A number of authors in the past have considered possible contributions to physical processes in QCD by studying momentum space Green's functions\cite{grossandrei,appelquistshankar,itep}.
Such computations, remarkably, are infrared finite, and arguably might represent distinguishable effects.  But there are reasons
to be skeptical.  Most strikingly, in the cases which have been studied, these instanton contributions fall very rapidly with momentum.  It is not clear that, e.g. in large $N$,
the leading contributions to the corresponding quantities won't swamp the semiclassical contributions.

In this note, we resolve this question by considering a set of correlation functions which receive no perturbative contribution but do receive
contributions from instantons.  They receive other non-perturbative contributions as well, as readily
seen in the large $N$ limit.  By considering the structure of the Operator Product Expansion (OPE), we can
show that, for particular numbers of flavors and colors, instantons provide the most singular contribution to these Green's functions
at short distances, while for others
they are subdominant.  In the former cases,
as we demonstrate self-consistently, a complete semiclassical computation is possible.  There is a systematic expansion of the most singular
  part of the coefficient function.  The leading instanton contribution to the singularity is infrared finite, and
corrections can be calculated systematically in powers of $\alpha_s$.  Dilute instanton gas corrections are infrared divergent, and can be organized as corrections
to the matrix elements of higher dimension operators, as well as less singular corrections to coefficient functions.

There is also a satisfying connection to large N arguments.  In the large $N$ limit, the coordinate space instanton contributions
to these correlators are always infrared divergent, and the
momentum space expression, while formally finite, is exponentially small compared to the leading non-perturbative contributions.  These latter can be determined
systematically in terms of $\langle \bar q q\rangle$.  The computation provides an explicit realization of Witten's assertion that instanton effects are exponentially
suppressed in large $N$\cite{wittencpn}.
A similar analysis demonstrates that in $e^+ e^-$ annihilation and deep inelastic scattering, instanton effects at large momentum transfers are highly subleading,
and cannot be discussed in a systematic fashion (independent of the number of flavors and colors).  As we will also explain, our observations may provide a
test of lattice computations which attempt to penetrate the chiral region.

The rest of this paper is organized as follows.  In the next section, we introduce the maximally chirality violating Green's functions which interest us.  We obtain
the condition that the instanton contributions are finite and show that, in such cases, there is always a power law singularity at short distances.  We explain
how to organize the computation in the language of the operator product expansion.  In the cases that the instanton contribution is infrared divergent, we show
that the infrared divergence corresponds to a correction to the matrix elements of operators already appearing in the perturbative OPE, as well as non-singular corrections
to the unit operator and operators of higher dimension.  In the case where the result is infrared finite, we demonstrate that the result represents a {\it calculable}, singular
contribution to the coefficient of the unit operator due to instantons.

In section \ref{dilutegas}, we discuss corrections and establish the nature
of the semiclassical expansion.  We show that perturbative corrections about the instanton yield corrections controlled by
$\alpha_s(x)$, and are systematically small at short distances.   The dilute instanton gas corrections are infrared divergent, and can again
be understood as corrections to matrix elements and non-singular corrections to OPE expansion coefficients.  In section \ref{largen},
we consider the large $N$ approximation, showing that one can systematically calculate the Green's functions in terms of
$\langle \bar q q \rangle$, and that the leading large $N$ contributions in momentum space are exponentially large compared
to the instanton contribution.  In section \ref{nfequalsnc}, we consider the case of $N_f = N$, which is the borderline
for calculability.  Here it is possible for instanton effects to dominate, but only by powers of logarithms.  We compute the power for
two interesting cases.  In section \ref{epluseminus}, we consider the application
of these observations to $e^+ e^-$ annihilation and similar processes, demonstrating that the instanton effects are subleading, and cannot be organized
as part of any systematic semiclassical expansion.  In section \ref{lattice}, we discuss applications to a possible test of lattice gauge theory
computations.  Finally, in our concluding section, we outline two directions for further work:  a study of these questions in $CP^N$ models, and a detailed
analysis of the lattice gauge theory problem.

\section{Instantons and Maximally Chirality Violating Green's Functions}
\label{mcg}

Consider a theory with $N$ colors and $N_f$ {\it massless} flavors, with $N_f$ small enough that the theory is asymptotically
free.  Label the flavors $q_f,\bar q_f$.   In an asymptotically free theory, we might hope that the short distance behavior of correlation functions can be systematically
analyzed.  The (gauge-invariant)  Green's function
\beq
G(x_i)=\langle \prod_{f=1}^{N_f} \bar q_f(x_f) q_f(x_f) \rangle
\eeq
vanishes in perturbation theory, and receives a contribution from
a single instanton.  For simplicity, take one set of points, say
$x_1,\dots, x_A = x$, and the other, $x_{A+1},\dots x_{N_f} =0$, where  $N_f = 2A$ or $N_f = 2A+1$, and call
\beq
{\cal O}_1(x) = \prod_{f=1}^A \bar q_f(x) q_f(x) ~~~~ {\cal O}_2(x)  =\prod_{f=A+1}^{N_f} \bar q_f(x) q_f(x).
\eeq
We study the two point function
\beq
\Delta(x) = \langle {\cal O}_1(x) {\cal O}_2(0) \rangle.
\eeq

We are interested in the behavior of $\Delta(x)$ for small $x$.  If the correlation function is singular, we should
be able to understand the singularities in terms of the operator product expansion.
As for any pair of operators, the OPE of  ${\cal O}_1(x),~{\cal O}_2(x)$ takes the form:
\beq
{\cal O}_1(x) {\cal O}_2(0) \approx \sum C_{12}^n(x){\cal O}_n(0)
\label{generalope}
\eeq
The coefficient functions, in an asymptotically free theory like QCD, can be calculated in a short distance expansion.  One expects that these include both perturbative
and non-perturbative contributions.  In the past, there has been much discussion of possible non-perturbative contributions to
OPE's\cite{appelquistshankar,grossandrei,itep}.
In general, this is a subtle problem.  If there are perturbative contributions to a coefficient, any non-perturbative contribution which
one might hope to isolate will be exponentially small.  In the case of $\Delta(x)$, there are no perturbative
contributions.
One can ask:  are any of the coefficient
functions dominated by instantons?

In perturbation theory, the theory respects an axial $U(1)$ symmetry, as a result of which the lowest dimension operator allowed on the right hand side
of eqn. \ref{generalope} is simply
$ {\cal O}_{3N_f}(0)={\cal O}_1(0) {\cal O}_2(0)$, with coefficient $C^{3N_f}(x)$.
$C^{3N_f}(x)$ starts out as a constant, and receives corrections in  powers of $\alpha_s(x)$ (we will discuss these in section \ref{nfequalsnc}).
More precisely, in perturbation theory, various operators of the same dimension as ${\cal O}_{3N_f}$ appear on the right hand side, multiplied
by powers of $\alpha_s(x)$; we will label these ${\cal O}_{3N_f}^a(0)$, and their coefficients $C^{3N_f}_a(x) $.  These will be considered
more explicitly in section \ref{nfequalsnc}.
Non-perturbatively, and in particular at the level of a single instanton, the {\it unit operator} can appear on the right hand side of this expression.
In other words,
\beq
{\cal O}_1(x) {\cal O}_2(0) \sim C^{(0)}(x) + C^{3N_f}_a(x)  {\cal O}_{3N_f}^a(0) + {\rm ~higher~dimension}.
\label{basicope}
\eeq
Assuming $C^{(0)}(x)$ is dominated by a well-defined instanton computation, we can determine its behavior on dimensional grounds (up to logarithmic
variation associated with the perturbative anomalous dimensions of ${\cal O}_1$ and ${\cal O}_2$):
\beq
C^{(0)} = \Lambda^{b_0} x^{-3N_f + b_0},
\eeq
where $b_0-3N_f={11\over 3}(N-N_f) $. So $C^{(0)}(x)$ exhibits a power-law singularity at short distances if $N_f>N$,  is non-singular if $N_f<N$,
and possesses a logarithmic singularity for $N_f = N$.
We will see that in the first case, the instanton computation is infrared finite, and there is a systematic expansion of the Green's function
in powers of $e^{-{8 \pi^2 \over g^2(x)}}$ and $g^2(x)$; in the second, the instanton computation is infrared divergent, and can be considered
to correct the matrix elements of ${\cal O}_{3N_f}$. The $N_f=N$ case requires more careful analysis.
Correspondingly, there are three behaviors for the correlation function.  For $N_f > N$,
\beq
\langle {\cal O}_1(x) {\cal O}_2(0) \rangle \approx C (\Lambda^{b_0} x^{-3N_f + b_0} + {\rm ~non~singular})
\eeq
where the coefficient $C$ is calculable, but the non-singular terms are not.  For
$N_f  < N$
\beq
\langle {\cal O}_1(x) {\cal O}_2(0) \rangle \approx D (\Lambda^{3N_f} \log(x \Lambda) + {\rm ~non~singular})
\eeq
where the coefficient, $D$, is {\it not calculable}.
For $N_f  = N$, the effects of instantons and other non-perturbative effects are comparable, and a more detailed analysis
is required.  This case, of course, is particularly interesting in real QCD, and will be the subject of section \ref{nfequalsnc}.

Note, most importantly, that in what we refer to as the non-calculable cases, the contributions of the operator
${\cal O}_1(0) {\cal O}_2(0)$ fall off more slowly than the instanton contributions, so, for example, the contributions
to this operator associated with dynamical breaking of the $SU(N_f)$ chiral symmetry are more important
than those calculated in the semiclassical approximation.

Let's consider the instanton computation in more detail.
In this section, we will examine the structure of the relevant integrals in order to determine
their behavior with $x$; particular cases important for applications
will be evaluated in more detail in section \ref{lattice} and still more thoroughly in \cite{weitaolarrylattice}.
The fermion zero modes have the form:
\beq
q_{\alpha}^{i} = {\rho {\sqrt{2} \over \pi}}{\delta^i_\alpha \over [(x-x_0)^2 + \rho^2]^{3/2}},
\eeq
where $\alpha$ is a two component spinor index and $i$ is the gauge index.
So the instanton contribution to $\Delta$ is
\beq
\Delta(x) = C\int d^4 x_0 d\rho {(\Lambda \rho)^{{11\over3} N -{2\over3} N_f} \rho^{3N_f -5} \over [(x-x_0)^2 + \rho^2]^{3A}[x_0^2 + \rho^2]^{3(N_f-A)}}
\eeq
where $C$ is a constant obtained from the non-zero modes in the instanton background, and $x_0$ and $\rho$ are the translational and rotational
collective coordinates, respectively.  The integral over $x_0$ may be performed by introducing Feynman parameters, yielding
\beq
\Delta(x) = C^\prime\int d\alpha [\alpha^{3A-1} (1-\alpha)^{3(N_f -A) -1}]  d\rho  {(\Lambda \rho)^{{11\over 3} N -{2\over 3} N_f} \rho^{3N_f -5} \over [x^2\alpha(1-\alpha) + \rho^2]^{3N_f-2}}.
\eeq
For large $\rho$, this behaves as
\beq
\Delta \sim \int {d\rho\over \rho} \rho^{{11\over 3}(N-N_f)}.
\eeq
The integral converges for large $\rho$ if $N_f >N$, exhibits a power law divergence for $N_f <N$, and diverges logarithmically
for $N_f = N$.  It
is also free of ultraviolet divergences (which would appear at the endpoints of the $\alpha$ integration),
and thus it behaves as guessed above:  .\footnote{Note it is important that we divided the fields into two groups as we did; otherwise, there would be ultraviolet divergences
associated with the definition of the local operators; this point is discussed further in section \ref{otheropes}.}

In the infrared divergent cases, the divergent part is identical to the (similarly ill-defined)
instanton contribution to $ \langle {\cal O}_1(0) {\cal O}_2(0) \rangle$.  For $N_f <N$, the (cutoff)
integral is non-singular for small $x$, corresponding to non-singular corrections to the coefficients of operators appearing in the OPE.
For the case $N_f=N$,  the expression also has a logarithmic singularity for small $x$, indicating the appearance of the unit operator in the OPE, with
a coefficient function behaving as $\log(x)$.  It is necessary to define the operators appearing in these expressions at a scale $M$, and this introduces
a mass scale both into the matrix element and into the coefficient of the unit operator.  We will discuss the scaling with $M$ in more detail in
section \ref{nfequalsnc}.
Indeed,
the question of whether we can compute the Green's function reliably is now a delicate one.  The operators ${\cal O}_{3N_f}^{a}$ possess anomalous dimensions.  As a result, the coefficient
functions behave as powers of logarithms.  We see that the coefficient of the unit operator is proportional to a single power of a logarithm.  So at best we can hope that $C^{(0)}$
dominates by a power of a logarithm.  We will explore this possibility in section \ref{nfequalsnc}.

\section{The Dilute Instanton Gas}
\label{dilutegas}

For the correlation functions we have been studying, there are two types of corrections we need to consider.
 First, there are perturbative corrections around the instanton.  Examining the Feynman diagram expansion, these corrections are seen
 to be infrared finite, and so lead to corrections given by powers of $\alpha_s(x)$ to the coefficient of the unit operator.  So we have, for the leading
 singular coefficient (displaying the result in a fashion which stresses the nature of the expansion),
 \beq
 \Delta(x) = {C \over x^{3N_f}}\exp \left ({-\int_{g(M)}^{g(x)} dg^\prime {\gamma_1(g^\prime) +\gamma_2(g^\prime)\over \beta(g^\prime)}}\right ) e^{-{8 \pi^2 \over g^2(x)}} \left(1 + \sum_{n=1}^{\infty} c_n\alpha_s(x)^n\right).
 \label{deltaexpansion}
 \eeq
 Here the first exponential reflects the anomalous dimensions of ${\cal O}_1$ and ${\cal O}_2$.

 Now we ask whether there are corrections in powers of $e^{-{8 \pi^2 \over g^2}}$.
Configurations containing $n+1$
instantons and $n$ anti-instantons would be governed by $e^{-(2n+1){8 \pi^2 \over g^2}}$.  As is well-known, in QCD there is not a controlled dilute instanton gas approximation, and
we need to establish that such configurations do not some how swamp the calculable single-instanton contributions
we have described above.  The essential point of this section is that these dilute gas contributions to the Green's functions
either represent corrections to the operator matrix elements, or non-singular corrections to the OPE coefficients.
In other words, the complete expansion of $\Delta(x)$ is of the form of eqn. \ref{deltaexpansion}.

A configuration with two instantons and one anti-instanton, widely separated, illustrates the main points.  Specialize to the case $N=2,N_f=3$, and
study the correlation function $\langle {\cal O}_1(x) {\cal O}_2(0) \rangle$.   Each of the widely separated instantons or anti-instantons
has six fermion zero modes.  One can absorb these in various ways, leading to different types of contributions to the correlator:
\begin{enumerate}
\item  ``Disconnected Diagrams":  Here all of the zero modes connected to one of the instantons are absorbed by the operators in the correlation function;
the zero modes emerging from the other instanton and the anti-instanton are tied together (i.e. connected by propagators).  The result is then the same as the one-instanton result, times an expression proportional
to the space-time volume.  This is a correction to the matrix element of the unit operator (building the exponential $e^{-EVT}$ in the Euclidean
path integral).
\item ``Connected diagrams":  Here there are contractions of fermion fields connecting the both instantons and the anti-instanton together.  All such diagrams are
infrared divergent, and do not exhibit singular behavior for small $x$.  The leading divergences are $x$-independent, and correct the matrix elements of operators such as
${\cal O}_1(0) {\cal O}_2(0)$.  Subleading, non-singular $x$-dependent terms can correct the coefficient of the unit operator.
\end{enumerate}
Higher orders in the semiclassical expansion behave similarly.

This argument establishes that widely separated instanton-anti-instanton configurations do not correct the singular part of the coefficient of the
unit operator in the operator product expansion, and thus do not correct the singular behavior of $\Delta(x)$.  Since the dilute gas expansion is not
under systematic control in QCD, this can be viewed only as strong evidence that eqn. \ref{deltaexpansion} represents the complete
expression for $\Delta(x)$, not a proof.  It should be possible to test some of these questions in
$CP^N$ models\cite{weitaolarrycpn}.

\section{The Large N Limit}
\label{largen}

If $N$ is large, and $N_f$ fixed, the behavior of the correlation functions at short distances is readily understood using large $N$ methods.
Consider the case of two flavors, $\bar u$ and $\bar d$.  We are interested in the short distance behavior of:
\beq
\Delta(x) = \langle \bar u(x) u(x) \bar d(0) d(0) \rangle
\eeq
From the perturbative OPE, the right hand side includes a term:
\beq
\langle \bar u(0) u(0) \bar d(0) d(0)\rangle \left ({\alpha_s(x) \over \alpha_s(\mu_0) }\right )^{c/b_0}
\eeq
The matrix element factorizes in the large $N$ limit, and so can be expressed in terms of $\langle \bar q q \rangle$ (defined at the scale $\mu_0$).

There will, in general, be additional operators on the right hand side, including the unit operator.  We will comment on this possibility shortly.
In any case, the expression above, in momentum space, behaves, at large momentum, as (it is more convenient to study the Fourier transform of $\nabla^2 \Delta(x)$, which
is better behaved)
\beq
\label{largeN}
\nabla^2 \Delta(x) \vert_{FT} \propto {\langle \bar q q\rangle^2 \over p^2} \times ~{\rm powers~of~\log(p)}.
\eeq
Instantons give a finite contribution to the momentum space Green's functions, so these behave as
\beq
\delta \Delta_{inst}(x) \propto p^6\left({\Lambda \over p}\right )^{{11 \over  3}N-4/3};
\eeq
for fixed $\Lambda/p$ they are exponentially suppressed relative to (\ref{largeN}) .  If there are large $N$ contributions to the
unit operator in the OPE,
(as one might reasonably suspect that there are), these may be even larger at large $p$.
In any case, we have here a realization of the slogan that instanton contributions at large $N$ are exponentially suppressed\cite{wittencpn}.

\section{$N_f = N$}
\label{nfequalsnc}

We have seen that in the case $N_f=N$, the operators with dimension $3N_f$ and the unit operator appearing in
 eqn. \ref{basicope} in the OPE behave similarly with $x$, up to powers of logarithms.  The best we might hope, then,
is that the instanton wins (or loses) by a power of a logarithm.  To assess this, we need to compute the anomalous dimensions of the various operators
appearing in the OPE.  This is a straightforward one loop computation.  Consider, first, the case of $SU(2)$ with two flavors.  Taking as the basis
of dimension six operators
\beq
{\mathcal O}_1 = \bar u u~ \bar d d ~~~{\mathcal O}_2 = \bar u \sigma^{\mu \nu}  u ~\bar d \sigma_{\mu \nu} \bar d
\eeq
the matrix of anomalous dimensions is:
\beq
\Gamma = A \left(
                                                                 \begin{array}{cc}
                                                                   15/2 & -2 \\
                                                                   0 & 3/2 \\
                                                                 \end{array}
                                                               \right)
\eeq
where
\beq
A = {g^2 \over 16 \pi^2} {2 \over \epsilon}.
\eeq
$\Gamma$ has eigenvalues
\beq
\gamma_1 = {15 \over 2} A~~~\gamma_2 = {3 \over 2} A.
\eeq
Correspondingly, given that the $\beta$ function in this theory is $6$, at small $x$, relative to the unit operator contribution, $\ln(x)$, the contributions
of the first operator behaves as
$(\log x)^{15/12}$, while those of the second behave as $(\log x)^{1/4}$.  So it is necessary to choose Green's functions carefully so as to obtain just the contribution of the second eigenoperator.

In $SU(3)$, the matrix $\Gamma$ is $3\times 3$, and there are not simple analytic expressions for the
eigenvalues, but one finds that, again, there is one operator which diverges more
slowly at small $x$ than the contribution of the unit operator.  In order to isolate an instanton
 contribution, it is necessary to take a linear combination of operators in the
 Green's function, $\Delta(x)$, which projects on this.   In practice, say in a lattice computation, this will be quite challenging, to say the least.

\section{Application:  Correlation Functions for $e^+ e^-$ annihilation}
\label{epluseminus}

Given our experience with chirality violating Green's functions,
it is interesting to ask whether we can isolate instanton effects in physical processes.  This is a question which has been investigated in the past (see, for example,
\cite{grossandrei,appelquistshankar,itep}).  Consider, for example, $e^+ e^-$ annihilation (well below the $Z^0$ threshold).  The hadronic total cross section is
proportional to
\beq
T \langle J^{\mu}(x) J^{\nu}(y) \rangle,
\eeq
Fourier transformed to momentum $q$.   The challenge is to separate out instanton effects from other {\it perturbative}
and {\it non-perturbative} effects.  The framework we have developed here is useful for this.   Before performing any calculations,
we note the following.  In perturbation theory, the unit operator appears in the operator product expansion, with a coefficient function
which behaves as $1/x^6$.  Fourier transformed, this gives the familiar form:
\beq
\Pi^{\mu \nu}(q) = (q^2 g^{\mu \nu} - q^\mu q^\nu) \Pi(q^2)
\eeq
with the perturbative contributions to $\Pi$ behaving as powers of logarithms.
  There is no reason to think that there aren't power law corrections at
large $q^2$, proportional to powers of $\Lambda^2/q^2$.

Let's consider instanton contributions.  More precisely, we need to consider dilute gas contributions with net instanton number zero.
In general, these will be ir divergent in coordinate space, unless $b_0 < 3/2$, and only for $b_0=0$ are they as singular
  at short distances as the perturbative contributions.  The leading IR divergences can be thought of as corrections to the
coefficient function of the unit operator (and higher dimension operators).  In momentum space, they will fall rapidly with momentum, and thus cannot be distinguished
from other types of non-perturbative effects (along the lines of our large N argument above).

Taking the lessons from our earlier discussion, we can establish the result by dimensional analysis.  The $\rho$ integral, in momentum space, is
always convergent\cite{appelquistshankar}, so the result of an instanton-anti-instanton contribution is necessarily:
\beq
\Pi(q^2) = \Lambda^{2 b_0} q^{-2b_0}.
\eeq
To be larger than the perturbative contributions, we require, again, $b_0 < 0$, i.e. we need be outside the regime where the theory is asymptotically
free.  If one examines the actual amplitudes, one finds expressions in agreement with that above (the coordinate space expression is infrared
finite for $b_0 < 2$).  Of course, in large $N$, we have an exponential suppression.

So, at least in this context, it does not seem possible to isolate instanton contributions to a physical, high momentum cross section.
Similar remarks apply to deep inelastic scattering and processes in heavy quark physics.  To find computable physical quantities requires,
at the very least, a theory in which chirality operating operators are important observables.  We do not see an obvious application to realistic scattering
experiments.  However, as we describe in the next section, there are potentially important applications of these ideas in lattice gauge theory.

\section{Applications:  Lattice Gauge Theory in the Chiral Limit}
\label{lattice}

A potential application of these ideas lies in tests of lattice gauge theory simulations.  What we have just learned
is that, for suitable numbers of flavors and colors, maximally chirality violating correlation functions can be computed
systematically in a semiclassical approximation.  Given that the study of the chiral limit of lattice gauge theories
is subtle, while, at the same time, current analyses probe quite small quark masses\cite{numericallattice}, comparison of ``experiment" and ``theory" could
be useful.

The realistic case, for which a significant amount of data exists (i.e. for which there already exists a large ensemble
of gauge configurations), would be $N_f = N = 3$.  We have seen that this is the borderline
situation for the semiclassical analysis; at the same time, the large
$N$ analysis is not reliable.   At most, the instanton computation dominates by a small power of a logarithm.  Still, it would
be of interest to examine the data to see {\it if} the Green's function exhibits roughly logarithmic variation with $x$, of a suitable order
of magnitude.
The simplest case to study where a reliable computation is possible is the case
$N=2, N_f=3$.  While not nature, this should be accessible to detailed numerical study.   Here
\beq
\langle {\cal O}_1(x) {\cal O}_2(0) \rangle = C{\Lambda^{16/3} \over x^{11/3}} (1 +{\cal O}( \alpha_s(x))).
\eeq
where the numerical coefficient, $C$,  is presented below in a particular scheme.
The question, then, is whether, with current lattice technology, such a computation is feasible; as we now
explain, we believe that, while
such a calculation is challenging, given the lattice spacings and quark masses currently being explored,
the answer may well be yes.  A detailed analysis of the issues will appear in reference\cite{weitaolarrylattice}; here we content ourselves with some
rough estimates.

In practice, one would compute the correlation function, dividing out by the vacuum to vacuum amplitude ($\langle { \mathbb
 I} \rangle$).  Current lattice computations
achieve quite small lattice spacings and quark masses..
For non-zero quark mass, perturbing in the quark masses, there are more singular contributions to the correlation function than those due to instantons.  (The relevant parameters are
$m_q/\Lambda$ and $1/(\Lambda x)$, where $\Lambda$ is the QCD scale).
%

First, we estimate more carefully the size of instanton effects.
The needed functional determinants can be obtained from 't Hooft's original paper\cite{thooft}.  Here we will content ourselves with a brief summary
of special cases in the $\overline{MS}$ scheme; more general results, including lattice regulators, will appear in ref. \cite{weitaolarrylattice}.  For
the case $N_f=3, N=2$, we obtain:
\beq
\Delta(x) \approx 9 \times 10^3 \mu^{16/3} g^{-8} e^{-8 \pi^2 \over g^2(\mu)} x^{-11/3}
\label{deltasu2}
\eeq
while the logarithmically singular term, for $N_f =3, N=3$ is:
\beq
\Delta(x) = -1.0 \times 10^7 \mu^{9} g^{-12} e^{-8 \pi^2 \over g^2(\mu)} \log(M~x).
\label{deltasu3}
\eeq

To estimate the effects on the lattice size, we will take the smallest lattice spacing and quark masses considered by the MILC collaboration\cite{numericallattice};
roughly $a = (4 ~{\rm GeV})^{-1}$, and $m_u,m_d \approx 10-20 {\rm ~MeV}$.  We will consider measuring the correlation function at $x= (1.5 ~{\rm  GeV})^{-1}$ (by these scales,
we mean scales such that the $SU(2)$ gauge coupling has the value corresponding to the observed $SU(3)$ coupling at that scale).
The leading terms are independent of $x$ and quite sensitive to the cutoff, $a$; roughly
they give a contribution, in the case of three flavors:
\beq
\Delta = {m_u m_d m_s \over a^6}.
\label{cutoffdependent}
\eeq
Using the expression for $\Delta$, eqn. \ref{deltasu2}, this is overwhelmingly larger than the instanton contribution.
This problem can be avoided by choosing operators which are odd under parity (the parity symmetry in the case of staggered fermions is
discussed, for example, in \cite{golterman}).  In that case, the leading contribution arises from a four loop diagram.  This behaves as (a detailed
computation will be presented in \cite{weitaolarrylattice}):
%
\beq
\Delta = c\left ({\alpha_s \over 4 \pi} \right )^3  {m_u m_d m_s \over a^2 x^4}.
\eeq
and should be much {\it smaller} than the instanton contribution.

One then can ask whether the instanton contribution is large enough that it is realistic to hope to extract it.  We believe the answer is yes, but that the
computation is likely to be challenging.  A yardstick (fermistick?) for comparison is
\beq
\Delta^\prime = S_F^3 \sim \left ( {1 \over  \pi^2}\right )^3 \vert x \vert^{-9}.
\eeq
with some contraction of the indices on $S_F$.  This is roughly comparable in size to $\Delta$ for $x^{-1} = 1.5$ GeV.  So one might hope that the numerical
computation will not be so noisy as to mask the instanton contribution.  Similar statements apply to the $SU(3)$ computation (which, again, is subject to
greater uncertainties, but is clearly of interest).  Examining somewhat similar numerical computations reinforces our optimism that this is a challenging,
but tractable, computation\cite{ukqcd}.

\section{Other OPE's, and the Definition of the Composite Operators}
\label{otheropes}

In section \ref{mcg}, we studied Green's functions for operators where we split the $2N_f$ fermion fields more or less evenly.   If we do not do this, in general, for
$N_f  > N+1$, we encounter {\it ultraviolet} divergences.  These can, again, be usefully understood in the language of the operator product expansion; if we define
the local operators by point splitting, then, under some circumstances, the instanton contribution to the point split operator diverges as the splitting tends to
zero.

Consider the $N=2, N_f=3$ case.  Here we would expect that the OPE of the operators $\bar u u(x) \bar d d(0)$ includes $\bar s^* s^*$ (in two
component notation).  The question is whether the OPE coefficient is singular.  It is easy to check that the three point function,
\beq
\langle \bar u(x) u(x) \bar d(0) d(0) \bar s(y) s(y) \rangle
\label{threedifferentpoints}
\eeq
in the limit $\vert x \vert^2 \ll \vert y \vert^2$, is non-singular.  So the corresponding OPE coefficient, while non-zero, is also non-singular as $x \rightarrow 0$.
More precisely, the OPE of pairs $\bar q_i q_i$ includes terms:
\beq
\bar u(x) u(x) \bar d(0) d(0) \sim \bar u(0) u(0) \bar d(0) d(0) + f(x) \bar s^*(0) s^*(0) + \dots
\label{chiralope}
\eeq
The statement that the Green's function of eqn. \ref{threedifferentpoints} is non-singular for small $x$ is the statement  that the coefficient function, $f(x)$, is non-singular
for small $x$.  The leading term in the Green's function, indeed, is precisely the matrix element of the first operator in eqn. \ref{chiralope}.

As in the case of the unit operator we encountered earlier, if we {\it increase} the number of flavors, for fixed $N$, the analog of $f(x)$ becomes singular.  For example,
for $N=2,N_f=4$, one can study the OPE:
\beq
\bar u(x) u(x) \bar d(x) d(x) \bar s(0) s(0) \sim \bar u(0) u(0) \bar d(0) d(0) \bar s(0) s(0) + f(x) \bar t^*(0) t^*(0) + \dots
\eeq
where now $f(x) \Lambda^{-14/3}\sim x^{-4/3}$.  Again, the coefficient function can be calculated using instantons.

So, for example, for four flavors in $SU(2)$, if we study the Green's function:
\beq
\langle \bar u(x) u(x) \bar d(x) d(x) \bar s(x+\epsilon) s(x+\epsilon) \bar t(0) t(0)\rangle
\eeq
and simply set $\epsilon=0$, we obtain an ultraviolet divergence, associated with the singularity of the operator product expansion discussed above.

\section{Conclusions and Directions for Further Research}

We have seen that semiclassical methods are sometimes applicable to QCD, including cases
with realistic numbers of flavors and colors; we have also seen that for many questions, as anticipated
long ago, instanton methods are not useful\cite{wittencpn}.  There are Green's functions which can be computed
systematically using instanton methods if $N_f>N$.  The case $N_f=N$ is borderline, in which, at most,
the instanton dominates at short distances by a power of a logarithm.  For $N \gg N$, we have seen
an explicit realization of the expectation that instanton effects are exponentially at small $N$.  On the other hand,
the most interesting case of $N_f=N =3$ exhibits no such suppression.

An interesting arena in which to analyze all of these questions is provided by the $CP^N$ models.  For such theories, an exact large $N$ solution is available, and it
should be possible to test a number of the ideas developed here.  It would, however, be especially interesting to attempt to study analogous Green's functions at finite $N$.
These issues will be discussed in a subsequent publication\cite{weitaolarrycpn}.

The most immediate application of these results is likely to be in lattice gauge theory.  Given the state of the art of current lattice computations, which probe quite short distances
and small quark masses,
one might hope to extract some of the Green's functions discussed here, comparing not only their qualitative behavior but their precise numerical values.
Results for the relevant Green's functions, including the effects of finite quark mass, will be presented in \cite{weitaolarrylattice}.  Successful agreement would provide
strong evidence for the accuracy of such lattice computations, and make important assertions, such as the fact that $m_u \ne 0$, all the more convincing.

\noindent
{\bf  Acknowledgements}

\noindent
Conversations with Tom Appelquist,
Tom Banks, Michael Peskin, Steve Shenker and Leonard Susskind are gratefully acknowledged as are comments by Zohar Komargodski and Lorenzo Ubaldi.   Remarks by Claude Bernard on staggered fermions and computational lattice issues were quite helpful.  This work supported in part by the U.S. Department of Energy.
M. Dine thanks Stanford University and the Stanford Institute for Theoretical Physics for a visiting faculty appointment while much of this work was performed.

\end{document}